\documentstyle[11pt]{article}

\newcommand{\xn}{x_{n}}

\newcommand{\kim}{ k_{1}^{\mu}}                                      
\newcommand{\kom}{ k_{0}^{\mu}}

\newcommand{\kn}{ k_{n}}

\newcommand{\km}{ k_{m}}

\newcommand{\ko}{ k_{0}}

\newcommand{\kon}{ k_{0}^{\nu}}

\newcommand{\lpp} {e^{i \int _{c} \alpha (s) 
k(s) \partial _{z} y(z+s) ds +ik_{0}y(z)}}   
\newcommand{\lppX} {e^{i \int _{c} \alpha (s) 
k(s) \partial _{z} X(z+s) ds +ik_{0}X(z)}}

\newcommand{\gvk}{ e^{i\sum _{n \ge 0 }k_{n}Y_{n}(z)}}

\newcommand{\dsi}{\frac{\partial}{\partial x_{1}}}

\newcommand{\dsq}{\frac{\partial }{\partial x_{n+m}}}           
\newcommand{\dst}{\frac{\partial }{\partial x_{2}}}

\newcommand{\dds}{\frac{\delta}{\delta \sigma}}                      
\newcommand{\ddS}{\frac{\delta}{\delta \Sigma}}

\newcommand{\dsnm}{\frac{\partial ^{2}}{\partial x_{n}\partial x_{m}}}

\newcommand{\p}{\partial}                                           
                                           
\newcommand{\pp}{\partial ^{2}}

\newcommand{\al}{\alpha }                                             
 
\newcommand{\tY}{\tilde Y}                                 
\newcommand{\lan}{\langle}
\newcommand{\ran}{\rangle}

\newcommand{\la}{\mbox{$ \lambda $}} 
\newcommand{\be}{\begin{equation}}
\newcommand{\br}{\begin{eqnarray}}
\newcommand{\ee}{\end{equation}} 
\newcommand{\er}{\end{eqnarray}}

\newcommand{\ppp}{\mbox {$ \partial ^{3}$}}

\begin{document}
\renewcommand{\theequation}{\thesection.\arabic{equation}}

\title{
\hfill\parbox{4cm}{\normalsize IMSC/2004/11/38\\
                               hep-th/0412033}\\        
\vspace{2cm}
Loop Variables and the (Free) Open String in a Curved Background. 
\author{B. Sathiapalan\\ {\em Institute of Mathematical Sciences}\\
{\em Taramani}\\{\em Chennai, India 600113}\\ bala@imsc.res.in}}           
\maketitle     

\begin{abstract} 
Using the loop variable formalism as applied to a sigma model in curved
target space, we give a systematic method for  writing down gauge and 
generally covariant 
equations of motion for the
modes of the free open string in curved space. 
The equations are obtained by covariantizing the flat space equation and 
then demanding gauge invariance, which introduces additional curvature
couplings.  
As an illustration of the procedure, the spin two case is worked out explicitly.

\end{abstract}

\newpage
\section{Introduction}

The loop variable approach is a proposal for writing down gauge invariant
equations of motion for the string. It has been applied to both the open 
and closed bosonic string theories and interacting gauge invariant 
equations have been written down. As it is based on the sigma model 
renormalization group \cite{L,CDMP,AS,FT}
 there is the possibility of addressing the issue of
manifest background independence in string theory \cite{EW}. We
 thus try and generalize the loop variable construction to curved
space time backgrounds. As a first step we do this for the free open string.
We give a prescription for writing down 
 gauge invariant and generally covariant equations of motion for
all the modes of the string generalizing the results of \cite{BSLV} to curved
space. We first do a straightforward covariantization of the
open string equations of motion. This violates gauge invariance. We show
that gauge invariance can be restored by adding curvature couplings.
These can be obtained systematically. We give the general procedure and
illustrate it with the simplest non-trivial case, which is spin-2. The spin-2
equations in curved space 
can also be  obtained by Kaluza-Klein reduction of Einstein gravity in 
one higher dimension \cite{CADS}. The loop variable method however can be just as easily applied
to arbitrarily high spin fields.   

 In principle one should be able to derive all the couplings to the curvature tensor
by including the  interaction of the bulk and boundary in the sigma model approach and if
done correctly, the resulting equations should be gauge invariant.
We have not done this in this paper. We have obtained the couplings 
just by requiring gauge invariance.
Assuming that there is a unique gauge invariant and generally covariant 
equation, we should get the same answer both ways. However it would be interesting
to explicitly check this. 

In the loop variable approach the self-interactions can usually be obtained by 
thickening the loop to a band \cite{BSREV,BSCP,BSC}. We expect this will be the case here also,
although we have not addressed this issue in this paper.

The problems that one faces when one tries to write higher spin equations
in curved space has been the subject of much investigation and 
solutions have also been given \cite{AD,BBB,BBD,BK,FV,MV}. It would be interesting to find
the relation between these results and the present work which is in the context of
 string theory.

This paper is organized as follows. In Section 2 we briefly 
review the sigma model
in curved target space and the Riemann Normal Coordinate expansion. In 
Section 3 we apply it to the loop variable to get covariant equations
of motion and explain how to modify them to make them gauge invariant.
  Section 4 contains conclusions.

\section{Sigma Model in Curved Space}

\subsection{Riemann Normal Coordinates}

In this section we remind the reader about Normal coordinates \cite{P}.
Let $x^\mu$ be a coordinate system with the notation 
$x^\mu (P) = x^\mu$ where $P$ denotes a generic point of the manifold,
and let $x^\mu _0$ be the coordinates
of a point $P_0$ in the manifold, at which we define Riemann Normal 
Coordinates (RNC). In the RNC the location of a generic point
$P$ is described  by a set of real numbers $y^\mu$ that are defined
 by the following:
Consider the geodesic $x^\mu(s)$ that goes through $P_0$ as well as $P$. 
The parameter $s$ can be chosen to be equal to
the distance along the geodesic. So $x^\mu (0) = x^\mu _0$.
Let the unit
tangent vector to this geodesic at $P_0$ be $\xi ^\mu$. Thus
\be   \label{xi}
\xi ^\mu ~=~ {dx^\mu \over ds}|_{x_0}
\ee
 If  $s$ is the distance to
point $P$ then $y^\mu = s~\xi ^\mu$ defines coordinates of the point $P$.    
While the tangent vector to the geodesic at $P_0$ is a geometric object
the components $\xi ^\mu$ defined above are with respect to the original $x$- 
coordinate axes.  One can equally well choose any other basis in the tangent
 space at $x_0$ to define the components.
 Thus the RNC  and the vectors $\xi ^\mu$
are independent of our original coordinate system $x^\mu$.  
But $x$ can be expressed as a (in general
nonlinear) function of $y$, $x(y)$, with $x(0) = x_0$.
$\xi$ and hence $y$ is a contravariant vector defined at the point $P_0$,
and so if we change our basis in the tangent space the vectors transform
 contravariantly.
In particular if we use (\ref{xi}) to define the components,
then the change in coordinate from $x$ to $x'$, will change
$y'^\mu = ({\p x'^\mu \over \p x^\nu})|_{x_0}~y^\nu ~=~ a^\mu _\nu y^\nu $.
 Here
the matrix $a^\mu _\nu$ are a set of numbers.

An immediate consequence of all this is the following. Consider the 
Taylor expansion of a tensor field at the point $x(y)$ about it's value at the point 
$x(y=0)=x_0$  :

\be
W^{\mu \nu ..}(x (y)) = W^{\mu \nu ..}(x (0)) + 
y^\rho {\p  W^{\mu \nu ..}(x (0))\over \p y^\rho}|_{x(0)}
+ {1\over 2!}y^\rho y^\sigma {\p ^2 W^{\mu \nu ..}(x(0)\over \p y^\rho \p y^\sigma}|_{x_0}+...
\ee

This expansion is valid in any coordinate system but if we let $x$ be a RNC
we have some simplification. 
Let  $x ^\mu = x_0^\mu + y^\mu$
exactly. In this coordinate system the geodesics at $P_0$ 
look like straight lines.
Thus $\Gamma ^\mu _{\nu \rho}(x_0)=0$, although the derivatives of $\Gamma$
are not zero. The derivatives obey 
\be \label{RNC}
\partial _{(\mu _1}\Gamma ^\nu _{\rho \sigma )} =
\partial _{(\mu _1 \mu _2}\Gamma ^\nu _{\rho \sigma )} =
\partial _{(\mu _1 \mu _2 ...\mu _r}\Gamma ^\nu _{\rho \sigma )} =~0
\ee

Thus for instance using $\Gamma (x_0)=0$ and (\ref{RNC}) and also,
\be
R^\nu _{~\rho \mu \sigma }(x_0) = 
\partial _\mu \Gamma ^\nu _{\rho \sigma} -
\partial _\sigma \Gamma ^\nu _{\rho \mu}
\ee 
 we get 
\be
\partial _\mu \Gamma _{\rho \sigma}^{\nu}(x_0) = {1\over 3} 
(R^\nu _{~\rho \mu \sigma}(x_0) + R^\nu _{~\sigma \mu \rho}(x_0))
\ee

Using these results one can write down relations of the following type:
\[
\partial _\mu W_\alpha (x_0) = D_\mu W_\alpha (x_0) + 
\Gamma _{\alpha \mu}^\beta W_\beta (x_0)
 = D_\mu W_\alpha (x_0) 
\]
\[
\partial _\nu \partial _\mu W_\alpha (x_0) = (\partial _\nu 
(D_\mu W_\alpha (x_0) + \Gamma _{\alpha \mu}^\beta W_\beta (x_0))
~=~D_\nu D_\mu W_\alpha + (\partial _\nu 
\Gamma _{\alpha \mu}^\beta )W_\beta (x_0)
\]
\be \label{Vec}
~=~D_\nu D_\mu W_\alpha + {1\over 3}(R^\beta _{~\alpha \nu \mu} ~+
~R^\beta _{~\mu \nu \alpha })W_\beta
\ee
and more generally one can write a ``covariant'' Taylor expansion \cite{P}:

\[
W_{\al _1 ....\al _p}(x) 
= W_{\al _1 ....\al _p}(x_0) ~+~
W_{\al _1 ....\al _p , \mu}(x_0)y^\mu ~+~
\]
\[
{1\over 2!}\{W_{\al _1 ....\al _p ,\mu \nu}(x_0) 
~-~{1\over 3}
\sum _{k=1}^p R^\beta _{~\mu \al _k \nu}(x_0) 
W_{\al _1 ..\al _{k-1}\beta \al _{k+1}..\al _p}(x_0)\}
 y^\mu y^\nu
~+~
\]
\[   
{1\over 3!}\{W_{\al _1 ....\al _p ,\mu \nu \rho}(x_0) - 
\sum _{k=1}^p R^\beta _{~\mu \al _k \nu}(x_0) 
W_{\al _1 ..\al _{k-1}\beta \al _{k+1}..\al _p, \rho}(x_0)
\]
\be \label{Taylor}
-
{1\over 2}\sum _{k=1}^p R^\beta _{~\mu \al _k \nu ,\rho }(x_0) 
W_{\al _1 ..\al _{k-1}\beta \al _{k+1}..\al _p}(x_0)\}y^\mu y^\nu y^\rho +...
\ee

Thus in the RNC we see that each term in the Taylor expansion 
is manifestly a tensor (defined at $x_0$) and thus the sum is a tensor at the point
$x_0$. Thus we see that the Taylor expansion, while valid in any coordinate system, becomes
in the RNC, a sum of tensors. However the RHS is a tensor at $x_0$ and the LHS is a tensor
at $x$. Thus this equation (\ref{Taylor})
is not a covariant tensor equation. It holds only in the RNC.

\subsection{Sigma Model with Curved Target Space}

Following \cite{AGFM,DF} we can consider a sigma model in curved background, with bulk action:
\be
S[X ] = -{1\over 2}\int d^2 \sigma G_{\mu \nu}(X) \partial _\alpha X ^\mu (\sigma , \tau )
 \partial _\alpha X ^\nu (\sigma , \tau ) 
\ee

We use RNC coordinates $y^\mu$ and Taylor expand all the tensors.
 Thus $G_{\mu \nu} (X) = G_{\mu \nu}(x_0) +
O(y^\mu )$.  We neglect the higher order terms. As stated in the introduction 
these are responsible for
closed string - open string interactions beyond those that are obtained
by covariantizing derivatives. We are going to obtain these by requiring gauge invariance. 
We choose a RNC 
\be
X ^\mu (\sigma , \tau) = x_0^\mu + y ^\mu (\sigma , \tau )
\ee
where $x_0$ is a constant (independent of $\sigma, \tau$) and is the reference 
point of our RNC.
The kinetic term for $y^\mu$ is $G_{\mu \nu}(x_0) \p _\al y^\mu \p ^\al y^\nu$. 
Generalizing the flat case prescription \cite{BSLV}, we set for the coincident
propagator:
\be
<y^\mu (\sigma ^\al) y^\nu (\sigma ^\al)> = G^{\mu \nu}(x_0) \rho (\sigma ^\al)
\ee 
where $\rho$ is the Liouville mode of the world sheet. 
 
We now consider the loop variable (LV) 
 $\gvk$, obtained as usual by the Taylor expansion of
$\lpp$, where 
\be
Y = y + \al _1 \p _z y + \al _2 \pp _z y + \al _3 {\ppp _z y \over 2!}+...
\ee
 and $\al _n$ are the modes of $\al$. \footnote{These are all described in \cite{BSLV,BSREV}. 
We refer the reader to those papers. A short summary is given in the Appendix.}
Correspondingly, again
generalizing \cite{BSLV}, we have
\be
<Y^\mu (\sigma ^\al) Y^\nu (\sigma ^\al) > = G^{\mu \nu}(x_0) \Sigma (\sigma ^\al)
\ee
Apart from the presence of the background metric, there is no difference with the flat space
case. Thus the equations of motion are obtained by requiring

\be       \label{lv}
\ddS :\gvk : e^{\sum _{n,m \ge 0 } \kn .\km {1\over 2}(\dsnm - \dsq )\Sigma } =0
\ee
 Here $k_0.y = k_\mu y^\mu$ and all space time indices are contracted using
$G_{\mu \nu} (x_0)$. $k_{0\mu} $ will thus become $-i{\p \over \p y^\mu}$.
The change
\be
\kn \rightarrow \kn + \la _p k_{n-p} 
\ee
changes the loop variable (\ref{lv}) by a total derivative ${\p \over \p x_p}$
and leaves the equations obtained from it invariant. These are the 
gauge transformations.

\section{Application to Loop Variables: Covariant Equations of Motion}

We can now proceed in the usual way to get the equations of motion in terms
of loop variables \cite{BSLV}. The  difference with the flat space case
will show up when we map loop variables to space time fields since $k_0^\mu$
will be mapped to covariant derivatives  as described in the previous 
paragraph. Thus we will obtain a covariant equation 
in terms of tensors at the point $x_0$. Since $x_0$ is arbitrary, we can take
this as the (covariant) equations of motion that we are after. The second step
is to obtain the gauge transformation in terms of space time fields. 

\subsection{Covariantizing the Equations}
\setcounter{equation}{00}
We give below, the loop variable equation and the corresponding covariant
equation for each of the fields.   

{\bf Tachyon:}

The equation for the tachyon is:
\be
[k_0^\mu k_{0\mu} + k_0^5 k_0^5 ]  =0
\ee
In flat space (or in RNC) this becomes (on setting $(k_0^5)^2 = ~dim [V]-1$,
where $dim [V]$ is the world sheet dimension of the vertex operator 
corresponding to that mode \cite{BSLV}) :

\be
(-\p _\mu \p ^\mu  -1) \Phi (x) =0
\ee
The covariant equation is therefore:

\be
(-D_\mu D^\mu - 1)\Phi (x) =0
\ee 

There is no gauge transformation for the tachyon.

{\bf Photon:}
 
The loop variable equation for the vector (photon) is
\be
[k_0 ^\mu k_{0\mu}k_1^\nu - k_0^\mu k_{1\mu}k_0^\nu ]=0
\ee  

Where
\[
<\kim > = A^\mu .
\]

In RNC this is Maxwell's equation:
\be
\p ^\mu \p _{[\mu }A _{ \nu ] } (x)  =0
\ee
Using (\ref{Vec}) we get in curved space:
\be \label{phot}
D^\mu [\partial _\mu A_\nu - \partial _\nu A_\mu ]=0
\ee

The gauge transformation

\be \label{gt1}
k_1^\mu \rightarrow k_1^\mu + \la _1 k_0^\mu
\ee
leaves the LV equation invariant. Setting $<\la _1 > = \Lambda _1 (k_0)$,
gives us the gauge transformation 
$A_\mu \rightarrow A_\mu + \p _\mu \Lambda$. Since $\Lambda$ is a scalar,
this is unmodified in curved space and continues to be an invariance of
Maxwelll's equation in curved space (\ref{phot}).

{\bf Spin 2:}

There are three fields $<k_1^\mu k_1^\nu> = -S_{1,1}^{\mu \nu }$,
$<k_2 ^\mu > = -iS_2^\mu$ and $<k_2^5 > = S_2 ^5$. The last two are auxiliary
fields that are necessary for gauge invariance. We also have to make the
identifications $k_1^\mu k_1^5 = k_2^\mu k_0^5$ and $k_1^5k_1^5 = k_2^5 k_0^5$.
As explained in \cite{BSLV} this is necessary in order to reproduce
the string spectrum.
Furthermore $k_0^5 = 1$ because the spin two vertex operators have dimension 2.
 
The loop variable equation for spin-2 is obtained by equating the coefficients
of $\dsi Y^\mu \dsi Y^\nu , ~ \dst Y^\nu ,~ \dsi Y^5 \dsi Y^5$ in (\ref{lv}), 
to zero \cite{BSLV}:

\br  \label{sp2}
-(k_0^2 +(k_0^5)^2)k_1^\mu k_1^\nu + k_1^{(\mu} k_0^{\nu )} k_1.k_0 - \\
 k_0^\mu k_0^\nu
k_1.k_1 + k_1^{(\mu} k_0^{\nu )} k_1^5 k_0^5 -  k_1^5 k_1^5 k_0^\mu k_0^\nu & = & 0
 \nonumber \\ \nonumber \\
k_0^2 k_2^\mu - k_1^\mu k_1.k_0 + k_0^\mu k_1.k_1 - k_0^\mu k_2.k_0  
 &=& 0  \\
\nonumber \\
-(k_1.k_0)^2 + (k_0^2 +(k_0^5)^2)k_1.k_1 - 2k_1^5k_0^5 k_1.k_0 + 
k_0^2 k_1^5k_1^5 &=&0
\er

The LV equation is invariant under
\be \label{gt2}
k_2^\mu \rightarrow k_2^\mu + \la _1 k_1^\mu + \la _2 k_0^\mu ~~~,~~~
k_1^\mu \rightarrow k_1^\mu + \la _1 k_0^\mu
\ee

In RNC the equation for $S_{1,1}^{\mu \nu}$ is:
\be
(\p_\rho \p ^\rho - (k_0^5)^2) S_{1,1}^{\mu \nu} - \p _\rho \p ^{(\mu }
S_{1,1}^{\nu )\rho } + \p ^\mu \p ^\nu S_{1,1 \rho}^\rho + \p ^{(\mu} 
S_2^{\nu )} (k_0^5)^2 - \p ^\mu \p ^\nu S_2 k_0^5 =0
\ee

$k_0^5 =1$, but we leave it as it is for later convenience.

We can now ``covariantize'' this equation. Using a generalization of (\ref{Vec})
we get
\be
\p _\rho \p _\sigma S_{\mu \nu} =
 D_\rho D _\sigma S_{\mu \nu} 
+ {1\over 3} (R^\al _{~\mu \rho \sigma }+R^\al _{~\sigma \rho \mu})
 S_{\alpha \nu}+ 
{1\over 3} (R^\al _{~\nu \rho \sigma }+R^\al _{~\sigma \rho \nu}) S_{ \mu \al}
\ee

Making use of the above one finds the covariant equation:

\[
D^\rho D_\rho S_{\mu \nu} 
-(k_0^5)^2 S_{\mu \nu} 
\]
\[-[ D^\rho D_\nu S_{\mu \rho} + D^\rho D_\mu S_{\nu \rho} + {2\over 3} (R^{\al
~ \rho}_{~\mu ~ \nu} + R^{\al ~ \rho}_{~\nu ~ \mu})S_{\al \rho}]
\]
\be  \label{cov2}
+D_\mu D_\nu S_{1,1\rho}^{\rho} +(D_\mu S_{2\nu}+ D_\nu S_{2\mu})(k_0^5)^2
 - D_\mu D_\nu S_2 k_0^5 =0.
\ee

The gauge transformation (\ref{gt2}) in RNC reads
\be
\delta S_{1,1\mu \nu} = \p_{(\mu }\Lambda _{1,1\nu )} ~~~,~~~
\delta S_{2\mu} = \Lambda _{1,1\mu} + \p_\mu \Lambda _2 ~~~,~~~
\delta S_2 = \Lambda _2 k_0^5 .
\ee
where $<\la _1 k_1 ^\mu >= \Lambda _{1,1}^\mu$ and $<\la _2> = \Lambda _2$.

In manifestly covariant notation the gauge transformation of
 $S_{1,1}^{\mu \nu}$ is modified 
to
\be \label{GT2}
\delta S_{1,1\mu \nu} = D_{(\mu }\Lambda _{1,1\nu )} ~~~,~~~
\delta S_{2\mu} = \Lambda _{1,1\mu} + \p_\mu \Lambda _2 ~~~,~~~
\delta S_2 = \Lambda _2 k_0^5 .
\ee
while the others are unmodified.

It is easy to see that the equation (\ref{cov2}) is not invariant
under (\ref{GT2}). The reason is not hard to find. 
The gauge transformation of the space time derivative of a field
is usually taken  to be 
 the derivative of the gauge transformation of the field.
But this is not what the loop variable prescription gives. 
Unfortunately, the prescription of the loop variable is the one that 
keeps the equation gauge invariant!
As an example of this discrepancy consider the loop variable
expression $L$ 
\[
L = k_{0\rho} k_{1\mu}k_{1\nu}
\]
 which is mapped to $S$, the covariant 
derivative of $S_{1,1\mu \nu}$:
\be   \label{S}
S = D_\rho S_{1,1\mu \nu} 
\ee

Schematically we have the map
\[ 
{\cal M}~:~ L \rightarrow S
\]
The gauge variation of $L$ is $L^g$
\[
{\cal G}~:~ L \rightarrow 
L^g = L + \delta L = L+
k_{0\rho} \la _1 k_{0(\mu}k_{1\nu )}
\]
 which
leaves the spin 2 LV equation invariant.
However the change $\Delta S$ given by 
\[
{\cal M}~:~ L^g \rightarrow S^g  = S + \Delta S = 
S+ <k_{0\rho} \la _1 k_{0(\mu}k_{1\nu )}>=
\]
\be   \label{Del}
S+ D_\rho D_{(\mu} 
\Lambda _{1,1\nu )} + 
{1\over 3} (R^\al _{~\nu \rho \mu} + R^\al _{~\mu \rho \nu})\Lambda _{1,1\al}
\ee
 is {\em not} 
the (covariant) derivative of the gauge variation of $S_{1,1\mu \nu}$:
 i.e. using (\ref{GT2}) one expects that

\be  \label{ds}
{\cal G}~:~ S \rightarrow S+ \delta S = S+
 D_\rho  D_{(\mu}\Lambda _{1,1\nu )} 
\ee
which is not equal to $S^g$ calculated above, because of the extra curvature
coupling. In (\ref{cov2}) there are two derivatives of the field $S^{\mu \nu}$
and this problem is more acute - one gets not only curvature terms but 
derivatives of the curvature.

\subsection{Imposing Gauge Invariance}

One can contemplate some ways out of this conflict.
 If we forcibly set $ \delta S = \Delta S$ instead of using (\ref{ds})
we will have invariance of the equation. This would force us to vary
the background gravitational field under a gauge transformation of the open
 string fields. This would mean that the covariant derivative does not commute
with gauge transformations. 

Another possibility is to  change the action of $\cal M$ on $L$ (but not 
$\delta L$) so that
$S$ is modified to $S' = S+f$, but $\Delta S$ is the same. The we can set
$\delta S'= \delta S + \delta f = \Delta S$. 
If we can find such a tensor $f$ we are done.
This means we do not modify the gauge transformation 
(\ref{GT2}) but modify (\ref{S}). This would mean modifying the
space time equation of motion while keeping the loop equation
unchanged. The gauge variation of the new space time equation under the
conventional gauge variation will give all the additional curvature terms.
Let us illustrate this in the example above by finding $f$.  

Keeping in mind the fact that 
\be   \label{Lam}
\delta (S_{2\mu} - {D _\mu S_2 \over 2 k_0^5}) = \Lambda _{1,1\mu}
\ee
we modify $S$ to
\be   \label{S'}
S' = S+f= D_\rho S_{\mu \nu} + {1\over 3} (R^\al _{~\nu \rho \mu} + 
R^\al _{~\mu \rho \nu})
 (S_{2\al} - {D _\al S_2 \over 2 k_0^5}) \equiv F_{\rho \mu \nu} 
\ee

It is easy to see that $\delta S' = \Delta S$.
Thus we have found $f$.
Thus if we say that ${\cal M}: L \rightarrow S'$, we will find
that we reproduce  the variation (\ref{Del}) using the standard 
variations of the space time fields. In this way we have managed
 to reproduce the variation specified by the loop variable prescription
and this guarantees the invariance of the equation.

This solution to the problem is easily generalized. In every case 
one modifies
the map from loop variable expression to space time fields by adding some extra terms.
 These additional terms involve products
of curvature tensor and auxiliary fields. These terms are chosen so
that they have the required gauge variation. This is always possible 
because of relations of the form (\ref{Lam}). Such relations
 can always be found for all the gauge parameters. This
is because in a massive theory the gauge parameters are in one-to-one
correspondence with auxiliary fields. The fact that there is an inverse
power of $k_0^5$ in (\ref{Lam}) (which gives the mass),
 tells you that it is crucial
that we are dealing with a massive gauge field \footnote{ This is reminiscent
of the inverse power of the cosmological constant in higher spin
massless gauge theories described in \cite{MV}}. In string theory
this is always the case for the higher spin fields, so this is not a problem.
In the limit that the string tension goes to zero, the higher modes of 
the string become massless.
In this case one should obtain a  description of
massless higher spin gauge fields in curved space.

We will now apply the above technique to the case at hand, viz. (\ref{sp2}). 
Using the notation of the example above we let 
\be
L = 
k_{0\rho}k_{0\sigma} k_{1\mu}k_{1\nu}
\ee
We let
\[
{\cal M }~:~ L  (=k_{0\rho}k_{0\sigma} k_{1\mu}k_{1\nu})  
\rightarrow S' (=<k_{0\rho}k_{0\sigma} k_{1\mu}k_{1\nu}>)
\]
It is a tensor $F$ given by
\be 
S'=F_{\rho \sigma \mu \nu} =
D_\rho D _\sigma S_{\mu \nu} 
+ {1\over 3} (R^\al _{~\mu \rho \sigma }+R^\al _{~\sigma \rho \mu})
 S_{\alpha \nu}+ 
{1\over 3} (R^\al _{~\nu \rho \sigma }+R^\al _{~\sigma \rho \nu}) S_{ \mu \al}
+ f_{\rho \sigma \mu \nu}
\ee
(We have supressed the subscripts ``$1,1$'' in both $S$ and $\Lambda$ below.)
Then we evaluate
\be
{\cal M} : \delta L =
 <\la _1 k_{0\rho}k_{0\sigma}k_{0(\mu}k_{1\nu )}
= 
\Delta S = G_{\rho \sigma \mu \nu} 
\ee 

The exact expression for $G$ is given in the Appendix.
The prescription is to choose $f$ such that
\[
G_{\rho \sigma \mu \nu} = D_\rho D_\sigma D_{(\mu }\Lambda _{\nu )} 
+ {1\over 3} (R^\al _{~\mu \rho \sigma }+R^\al _{~\sigma \rho \mu})
 (D_\al \Lambda _\nu + D_\nu \Lambda _{\al})+
\]
\be
{1\over 3} (R^\al _{~\nu \rho \sigma }+R^\al _{~\sigma \rho \nu})
   (D_\al \Lambda _\mu + D_\mu \Lambda _{\al})+
\delta f_{\rho \sigma \mu \nu}
\ee

As an example, consider the term proportional to $D_\nu \Lambda _\al$.
We find that
\be
\delta f_{\rho \sigma \mu \nu} =
 [-{1\over 6}(R^\al _{~\mu \rho \sigma}
+R^\al _{~\sigma \rho \mu}) + {1\over 6}R^\al_{~\rho \sigma \mu} ]
D_\nu \Lambda _\al
\ee
This implies that we can take one of the terms in $f$ to be
\be
 [-{1\over 6}(R^\al _{~\mu \rho \sigma}
+R^\al _{~\sigma \rho \mu}) + {1\over 6}R^\al_{~\rho \sigma \mu} ]
D_\nu (S_{2\al}-D_\al {S_2\over 2k_0^5}) 
\ee
This is $f^1_{\rho \sigma \mu \nu}$ of the Appendix.

Thus we calculate $F_{\rho \sigma \mu \nu}$ which is  given in the Appendix.
$F_{\rho \mu \nu}$ is given in (\ref{S'}).
These in turn are determined by the 
 two tensors $f_{\rho \mu \nu}$ and $f_{\rho \sigma \mu \nu}$. 
The corresponding tensor modification 
for terms involving $S_{2\mu}$
and the scalar $S_2$ are zero.

In terms of these tensors the equations of motion are ($k_0^5 =1$):

\[
F^\rho_{~\rho \mu \nu} + (k_0^5)^2 S_{1,1}^{\mu \nu}  -F_{\rho (\nu \mu )} ^{~~~~~\rho} + 
F_{\mu \nu ~\rho}^{~~\rho} +
D_{(\mu}S_{2\nu )} (k_0^5)^2 - D_{(\mu}D_{\nu )} S_2 k_0^5 =0
\] 
\[
D^\rho D_\rho S_{2\mu} - F^\rho _{~\mu \rho} + F_{\mu ~\rho}^{~\rho} -
D_\mu D^\rho S_{2\rho} =0
\]
\be    \label{Spin2}
-F^{\rho \sigma}_{~~\rho \sigma}+F^{\rho ~\sigma}_{~\rho ~\sigma} -
(k_0^5)^2S_{1,1\rho}^\rho - 2 D_\rho S_2^\rho (k_0^5)^2 - D^\rho D_\rho S_2 
k_0^5 =0
\ee

By construction they are gauge invariant under 
\be 
\delta S_{1,1\mu \nu} = D_{(\mu }\Lambda _{1,1\nu )} ~~~,~~~
\delta S_{2\mu} = \Lambda _{1,1\mu} + \p_\mu \Lambda _2 ~~~,~~~
\delta S_2 = \Lambda _2 k_0^5 .
\ee

The same procedure can clearly be applied to spin 3 
\footnote{The LV equations for spin 3 are given in \cite{BSLV}}
and the higher modes also.

\section{Conclusions}

We have given a method of writing down gauge and generally covariant equations 
of motion for the higher spin modes of the  open bosonic string. Explicit equations for 
spin 2 are given (\ref{Spin2}) as an illustration of the method . The method
is a generalization of the flat space method of \cite{BSLV}. The main point
of departure, apart form the obvious one of covariantizing, is that the
map from loop varables to space time fields is modified. The modification is 
designed to reproduce the gauge transformation of the loop variable. This ensures
that the equations are gauge invariant. The modification involves adding
terms that involve inverse power of the mass of the higher string modes.
Since the mass is proportional to the string tension,
it would  be interesting to try and obtain the massless case as a limit
of vanishing string tension.

The loop variable method in the flat space case was easily generalized
to the interacting case by thickening the loop to a band \cite{BSREV}.
 If this holds in curved
space also then we have  a set of interacting higher spin massive fields
in curved space. 
This can then be treated as a step towards a proposal for a
 manifestly background independent formulation of string theory\cite{EW}.

{\bf Acknowledgements:} I would like to thank G. Date and S. Kalyana Rama  for
reading the manuscript and giving some useful comments.

\appendix
\section{Appendix: Calculation of $F$.}
\setcounter{equation}{00}

In this Appendix we calculate the tensor $F_{\rho \sigma \mu \nu}$.
The first step is to calculate $G_{\rho \sigma \mu \nu}$ defined
by
\[
G_{\rho \sigma \mu \nu} = <\la _1 k_{0\rho} k_{0\sigma} k_{0(\mu} k_{0\nu )}>
\]
We have defined this as $\p _\rho \p_\sigma \p_\mu \Lambda _\nu (0))$
($\p _\mu = {\p \over \p y^\mu}$)written in covariantised form. We use the Taylor expansion (\ref{Taylor})
to write
\[
\Lambda _\nu (y) = \Lambda _\nu (0) + y^\al D_\al \Lambda _\nu +
{ y^\al y^\beta \over 2!} (D_\al D_\beta \Lambda _\nu +
 {1\over 3} R^\sigma _{~\beta \al \nu} \Lambda _\sigma) +
\]  
\be
{y^\al y^\beta y^\gamma \over 3!} (D_\al D_\beta D_\gamma \Lambda _\nu 
-R^\sigma _{~\al \nu \beta }D_\gamma \Lambda _\sigma - {1\over 2}
D_\al R^\sigma _{~\beta \nu \gamma }\Lambda _\sigma)+...
\ee           

Acting on this with $\p_\rho \p _\sigma \p _\mu$ gives 
\[
{1\over 3!}( D_{~(\rho} D_\sigma D_{\mu )}\Lambda _\nu - 
R^\al _{~(\rho \nu \sigma} D_{\mu ) }\Lambda _\al -{1\over 2}
D_{(\rho} R^\al _{~\sigma \nu \mu )} \Lambda _\al)
\]
where the symmetrization is over the three indices $\rho , \sigma , \mu$.

We denote the three terms as I,II and III.

{\bf I:}

The first term has to brought into the form $D_\rho D_\sigma D_\mu$.
This can be achieved by using the commutation rules:
\[
[D_\rho ,D_\sigma ] W_{\mu } = R _{~\mu \sigma \rho}^\al W_\al
\]
\[
[D_\rho , D_\sigma ] W_{\mu \nu} = 
R _{~\mu \sigma \rho}^\al W_{\al \nu} +
R _{~\nu \sigma \rho}^\al W_{\mu \al}
\]

If we do this and symmetrize on $\mu ,\nu$ we get
\[
G^{I}_{\rho \sigma \mu \nu}=
D_\rho D_\sigma D_{(\mu}\Lambda _{\nu )} + 
{1\over 2}(D_\rho( R^\al_{~\nu \sigma \mu} + R^\al_{~\mu \sigma \nu}))\Lambda _\al
\]
\[
+ {1\over 6}(D_\mu R^\al_{~\nu \sigma \rho}+D_\nu R^\al_{~\mu \sigma \rho})\Lambda _\al
+{1\over 6}(D_\sigma (R^\al _{~\nu \rho \mu} + R^\al _{~\mu \rho \nu}))\Lambda _\al
\]
\[
+{1\over 2}(R^\al_{~\nu \sigma \mu} + R^\al _{~\mu \sigma \nu})D_\rho \Lambda _\al
+{1\over 6}(R^\al_{~\nu \rho \sigma}D_\mu \Lambda _\al + 
R^\al _{~\mu \rho \sigma}D_\nu \Lambda _\al)
\]
\[
+ {1\over 2}(R^\al _{~\nu \rho \mu}+R^\al _{~\mu \rho \nu})D_\sigma \Lambda _\al
+ {1\over 3} (R^\al _{~\sigma \rho \mu} + R^\al_{~\mu \rho \sigma})D_\al \Lambda _\nu
\]
\be
+ {1\over 3}(R^\al _{~\sigma \rho \nu} + R^\al _{~\nu \rho \sigma })
 D_\al \Lambda _\mu
\ee

{\bf II:}

The second term on symmetrization gives:
\[
G^{II}_{\rho \sigma \mu \nu} =
-{1\over 3!}[ (R^\al _{~\rho \nu \sigma } + 
R^\al _{~\sigma \nu \rho}) D_\mu \Lambda _\al
+(R^\al _{~\rho \mu \sigma } + 
R^\al _{~\sigma \mu \rho}) D_\nu \Lambda _\al
\]
\be 
+
(R^\al _{~\mu \nu \rho} + R^\al _{~\nu \mu \rho})D_\sigma \Lambda _\al +
(R^\al _{~\mu \nu \sigma} + R^\al _{~\nu \mu \sigma})D_\rho \Lambda _\al ]
\ee

{\bf III:}

The third term gives
\[
G^{III}_{\rho \sigma \mu \nu} = 
-{1\over 12} [
D_\rho R^\al _{~\mu \nu \sigma} +D_\rho R^\al _{~\nu \mu \sigma} +
D_\mu R^\al _{~\rho \nu \sigma} +D_\nu R^\al _{~\rho \mu \sigma} 
\]
\be
+
D_\mu R^\al _{~\sigma \nu \rho} + D_\nu R^\al _{~\sigma \mu \rho}+
D_\sigma R^\al _{~\mu \nu \rho} + D_\sigma R^\al _{~\nu \mu \rho} ]
\ee

Adding everything gives
\[
G_{\rho \sigma \mu \nu }= D_\rho D_\sigma D_{(\mu}\Lambda _{\nu )} +
{2\over 3} (R^\al _{~\nu \sigma \mu} + R^\al_{~\mu \sigma \nu})D_\rho \Lambda _\al +
{2\over 3} (R^\al _{~\nu \rho \mu} + R^\al_{~\mu \rho \nu})D_\sigma \Lambda _\al +
\]
\[
[{1\over 6} (R^\al_{~\nu \rho \sigma}  + R^\al _{~\rho  \sigma \nu} + 
R^\al _{~\sigma  \rho \nu})]D_\mu \Lambda _\al +
[{1\over 6} ( R^\al_{~\mu \rho \sigma}  + (R^\al _{~\rho \sigma \mu} + 
R^\al _{~\sigma \rho \mu })]D_\nu \Lambda _\al +
\]
\[
{1\over 3}(R^\al_{~\sigma \rho \mu} + R^\al _{~\mu \rho \sigma})D_\al \Lambda _\nu +
{1\over 3}(R^\al_{~\sigma \rho \nu} + R^\al _{~\nu \rho \sigma})D_\al \Lambda _\mu +
\]
\[
{7\over 12}(D_\rho R^\al _{~\mu \sigma \nu}+
 D_\rho R^\al _{~\nu \sigma \mu})\Lambda _\al +
{1\over 4}(D_\sigma R^\al _{~\mu \rho \nu}+
 D_\sigma R^\al _{~\nu \rho \mu})\Lambda _\al +
\]
\[
(-{1\over 4} D_\mu R^\al _{~\rho \nu \sigma} + 
{1\over 12}D_\mu R^\al _{~\sigma \nu \rho})\Lambda _\al +
\]
\be
(-{1\over 4} D_\nu R^\al _{~\rho \mu \sigma} + 
{1\over 12}D_\nu R^\al _{~\sigma \mu \rho})\Lambda _\al 
\ee

We have to set this equal to
\be
D_\rho D _\sigma \delta S_{\mu \nu} 
+ {1\over 3} (R^\al _{~\mu \rho \sigma }+R^\al _{~\sigma \rho \mu})
 \delta S_{\alpha \nu}+ 
{1\over 3} (R^\al _{~\nu \rho \sigma }+R^\al _{~\sigma \rho \nu}) \delta S_{ \mu \al}
+ \delta f_{\rho \sigma \mu \nu}
\ee

Comparing both sides we can write $f$ as the sum of five terms:
\[
f_{\rho \sigma \mu \nu} = \sum _{i=1}^5 f^i_{\rho \sigma \mu \nu} 
\]
with:  

\[
f^1_{\rho \sigma \mu \nu} = 
-{1\over 6}(R^\al_{~\mu \rho \sigma} + R^\al _{~\sigma \rho \mu} +
 R^\al _{~\rho \sigma \mu}) D_\nu (S_{2\al}- {D_\al S_2^5 \over 2k_0^5}) 
\]
\[
f^2_{\rho \sigma \mu \nu} = 
-{1\over 6}(R^\al_{~\nu \rho \sigma} + R^\al _{~\sigma \rho \nu} +
 R^\al _{~\rho \sigma \nu}) D_\mu (S_{2\al}- {D_\al S_2^5 \over 2k_0^5}) 
\]
\[
f^3_{\rho \sigma \mu \nu} = 
{2\over 3}(R^\al _{~\nu \rho \mu} + R^\al _{~\mu \rho \nu})D_\sigma (S_{2\al} -
 {D_\al S_2^5\over 2k_0^5})
\]
\[
f^4_{\rho \sigma \mu \nu} = 
{2\over 3}(R^\al _{~\nu \sigma \mu} + R^\al _{~\mu \sigma \nu})D_\rho 
(S_{2\al} - {D_\al S_2^5\over 2k_0^5})
\]
\[
f^5_{\rho \sigma \mu \nu} = 
[{7\over 12}(D_\rho R^\al _{~\mu \sigma \nu}+
 D_\rho R^\al _{~\nu \sigma \mu}) +
{1\over 4}(D_\sigma R^\al _{~\mu \rho \nu}+
 D_\sigma R^\al _{~\nu \rho \mu}) +
\]
\[
(-{1\over 4} D_\mu R^\al _{~\rho \nu \sigma} + 
{1\over 12}D_\mu R^\al _{~\sigma \nu \rho}) +
\]
\be
(-{1\over 4} D_\nu R^\al _{~\rho \mu \sigma} + 
{1\over 12}D_\nu R^\al _{~\sigma \mu \rho})](S_{2\al} - {D_\al S_2^5\over 2k_0^5})
\ee

Thus we finally get
\be
F_{\rho \sigma \mu \nu} = D_\rho D _\sigma S_{\mu \nu} 
+ {1\over 3} (R^\al _{~\mu \rho \sigma }+R^\al _{~\sigma \rho \mu})
 S_{\alpha \nu}+ 
{1\over 3} (R^\al _{~\nu \rho \sigma }+R^\al _{~\sigma \rho \nu}) S_{ \mu \al}
+ f_{\rho \sigma \mu \nu}
\ee

\section{Appendix: Loop Variables}
\setcounter{equation}{00}

We give a short summary of the loop variable approach here.
The basic idea introduced in \cite{BSLV} is to generalize the vertex operators 
of the sigma model by introducing
additional variables $\xn , n>0$ that, very roughly,
have the property that ${\p \over \p \xn }\approx \p _z ^n$.
They can be thought of as the modes of an einbein that make the loop variable
diffeomorphism invariant.

We thus consider the following loop variable:
\be   \label{84}
\lppX
\ee

It is understood that $Y_0=Y$.
$\al (s)$ is an einbein. Let us assume the following Laurent expansion:
\be
\al (s) ~=~ 1 ~+~ {\al _1 \over s}~+~{\al _2 \over s^2} ~+~ {\al _3 \over s^3}+...
\ee

Let us define 
\br
Y &~=~& X ~+~ \al _1 \p _z X ~+~ \al _2 \p_z^2 X ~+~
 \al _3 {\p _z ^3 X \over 2}~+~ ...~+~{\al _n \p _z^n X\over (n-1)!}~+~...\nonumber \\
&~=~& X ~+~ \sum _{n>0} \al _n \tY _n \\
Y_1 &~=~& \p _z X ~+~ \al _1 \p_z^2 X ~+~ \al _2 {\p _z ^3 X \over 2}~+~ ...~+~{\al _{n-1} \p _z^n X\over (n-1)!}~+~...\nonumber \\
...& & ...\nonumber \\
Y_m &~=~& {\p _z^m X\over (m-1)!} ~+~ \sum _{n > m}{\al _{n-m} \p _z^n X\over (n-1)!}\\
\er

If we define $\al _0 =1$ then the $>$ signs in the summations above can be replaced by $\ge$.

Using these equations one can write
\be
\lppX = \gvk
\ee

It is understood that $Y_0=Y$.

Let us now introduce $\xn$ by the following:
\be
\al (s) = \sum _{n\ge 0} \al _n s^{-n} = e^{\sum _{m\ge 0} s^{-m} x_m}
\ee

Thus 
\br
\al _1 &=& x_1  \nonumber \\
\al _2 &=& {x_1^2 \over 2} + x_2 \nonumber \\
\al _3 &=& {x_1^3 \over 3!} + x_1x_2 + x_3
\er

They satisfy the property,
\be
{\p \al _n \over \p x_m} = \al _{n-m} , ~~ n\ge m
\ee

Using this we see that 
\be
Y_n = {\p Y\over \p x_n}
\ee

Now we can impose conformal invariance by demanding $\dds =0$
where $\sigma$ is the Liouville mode : $<X(z) X(z)> = ~ln~ a +\sigma$
($a$ is the world sheet cutoff).
Except we will define  $\Sigma = \lan Y(z) Y(z) \ran$. This is
equal to the previous $\sigma$ in coordinates where $\al (s) =1$.
Thus we have for the coincident two point functions:
\br    \label{Sig}
\lan Y ~Y\ran &~=&~ \Sigma \nonumber \\
\lan Y_n ~Y\ran &~=&~ {1\over 2}{\p \Sigma \over \p x_n}   \nonumber \\
\lan Y_n ~Y_m \ran &~=&~  {1\over 2}({\pp \Sigma \over \p x_n \p x_m} - {\p \Sigma \over \p x_{n+m}})
\er

Using this the normal ordering gives:
\br   \label{LV}
\lppX &=& \gvk  \nonumber \\
&=& exp \{\ko ^2 \Sigma + \sum _{n >0} \kn .\ko  {\p \Sigma \over \p x_n} +  \nonumber \\
& & \sum _{n,m >0}\kn .\km {1\over 2}({\pp \Sigma \over \p x_n \p x_m} - {\p \Sigma \over \p x_{n+m}})\} \nonumber \\
& & :\gvk :
\er

We can now operate with  ${\delta \over \delta \Sigma}$ and set $\Sigma =0$. We will only give one sample variation here:

\[
{\delta \over \delta \Sigma} [
\kn .\km {1\over 2}({\pp \Sigma \over \p x_n \p x_m} - {\p \Sigma \over \p x_{n+m}})
] :e^{i\ko .Y}:
= :({1\over 2}i\kom i\kon Y_n^\mu Y_m ^\nu + i\kom  Y_{n+m}^\mu ) e^{i\ko .Y}:
\]  

One can thus collect all the coefficients of a particular vertex operator,
say $:Y_n ^\mu e^{i\ko .Y}:$, and this gives the free equation of motion. Note that
they never contain more than two space-time derivatives. 

We turn to the issue of gauge invariance. We have assumed that
$\al (s)$ is being integrated over, which is why we are allowed to integrate
by parts. This means that 
\be   \label{GT}
k(s) \to \la (s) k(s)
\ee
 is equivalent to $\al (s) \to \la (s) \al (s)$, which is clearly just a change
of an integration variable. Assuming the measure is invariant this does nothing
to the integral. The measure ${\cal D}\al (s)$ has been chosen to be $\prod _n d\xn $.
If we set $\la (s) = e^{\sum _m y_m s^{-m}}$ , then the gauge transformation (\ref{GT})
 is just a translation, 
$\xn \to \xn + y_n$ which leaves the measure invariant. Thus we conclude that
(\ref{GT}) gives the gauge transformation. 

If we expand $\la (s)$ in inverse powers of $s$
\[
\la (s) = \sum _n \la _n s^{-n}
\]
Then we can write (\ref{GT}) as 
\be    \label{GT1}
\kn \to \sum _{m=0}^{n} \la _m k_{n-m}
\ee

We set $\la _0 =1$.

In order to interpret these equations in terms of space-time fields 
we use (\ref{Free}). They have to be extended to include $\la$. Thus we
assume that the string wave-functional is also a functional of $\la (s)$. 
Thus we set
\br   \label{La}
\lan \la _1 \ran &~=&~ \Lambda _1 (\ko )\nonumber \\
\lan \la _1 \kim \ran &~=&~ \Lambda _{11}^\mu (\ko ) \nonumber \\
\lan \la _2 \ran &~=&~ \Lambda _2 (\ko )
\er

The gauge transformations (\ref{GT1}) thus become on mapping to
space time fields by evaluating
$\lan .. \ran$:

\br
A^\mu (\ko )~&\to &~ A^\mu (ko ) + \kom \Lambda _1 (\ko ) \nonumber \\
S^\mu _2 (\ko )~&\to &~ S^\mu _2 (ko ) + \kom \Lambda _2 (\ko ) + \Lambda ^\mu _{11} \nonumber \\
S_{11}^{\mu \nu} ~&\to &~ S^{\mu \nu}_{11} + k^{(\mu}_0 \Lambda _{11}^{\nu )}
\er  
These are more or less the canonical gauge transfomations for a massive spin two field.
\footnote{They become identical after we perform a dimensional reduction. This will be described later.}
Now it is known that the gauge transformation parameters of higher spin fields
obey a certain tracelessness condition \cite{F,SH}.  We will see this below also.

When one actually performs the
gauge transformations we find the following mechanism for gauge invariance.
It changes the
normal ordered loop variable by a total derivative in $\xn$ 
which doesn't affect the equation of motion.
More precisely the gauge variation of the loop variable is a term of the form
${d\over d\xn } [A(\Sigma ) B]$, where $B$ doesn't depend
on $\Sigma$. The coefficient of $\delta \Sigma$ is obtained as
\[
\int ~~\delta ({d\over d\xn } [A(\Sigma ) B]) =
\int ~~ ({d\over d\xn}( {\delta A\over \delta \Sigma } \delta \Sigma ) B +
  {\delta A\over \delta \Sigma } \delta \Sigma {dB\over d\xn })
\]
\[
=\int ~~[ - {\delta A\over \delta \Sigma } {dB\over d\xn } +
 {\delta A\over \delta \Sigma } {dB\over d\xn }]\delta \Sigma =0
\]
Here we have used an integration by parts.

Actually one finds on explicit calculation that the variation is not a total
derivative. This is because in deriving (\ref{Sig}) some identities
have been used. Thus only if we use those identities in the variation will
we be able to write the variation as a total derivative. However we do
not want to use them because we would like to leave $\Sigma$ unconstrained
when we vary. Thus constraints have to be imposed elsewhere. It can easily be
checked that the terms that have to be put to zero are all of the form
\be
\la _n \km . k_p ...
\ee   

where ... refers to any other factors of $k_m$ \cite{BSLV,BSLV'}. 
Thus all traces
of gauge parameters have to be set to zero. 
This thus explains the tracelessness
mentioned earlier.

In \cite{BSLV,BSLV'} some examples, namely spin-2 and spin-3
are explicitly worked out.

\end{document}